**Sleep and its relation to cognition and behavior in preschool-aged children of the general population: a systematic review**

Short title: Sleep, cognition and behavior in preschoolers

Authors: Eve Reynaud[1,2,3], Marie-Françoise Vecchierini MD, PhD[4], Barbara Heude PhD[1,2], Marie-Aline Charles MD, PhD[1,2], Sabine Plancoulaine MD, PhD[1,2]

Affiliations: [1] INSERM, UMR1153, Epidemiology and Statistics Sorbonne Paris Cité Research Center (CRESS), early ORigins of Child Health And Development Team (ORCHAD), Villejuif, F-94807 France;

[2] Paris-Descartes University, Paris, France;

[3] Ecole des Hautes Etudes en Santé Publique (EHESP), Rennes, F-35043 France

[4] Hôpital Hôtel Dieu, Centre du Sommeil et de la Vigilance, AP-HP, Paris, France; Sorbonne Paris Cité, EA 7320 VIFASOM, Université Paris Descartes, Paris, France.

Address correspondence to: Sabine Plancoulaine, INSERM U1153, Team 6 ORCHAD, 16 Avenue Paul Vaillant Couturier, 94807 Villejuif Cedex, France, [sabine.plancoulaine@inserm.fr], + 33 145-595-109

Conflict of interest: There are no financial or non-financial conflicts of interest. This research did not receive any specific grant from funding agencies in the public, commercial, or not-for-profit sectors

Author contributorship: Eve Reynaud and Sabine Plancoulaine developed the search strategy and conducted the selection of articles, data extraction and reporting. Eve Reynaud drafted the initial manuscript under the supervision of Sabine Plancoulaine. Marie-Françoise Vecchierini contributed to the search strategy, and gave guidance for the summary of the data. Barbara Heude and Marie-Aline Charles gave guidance on child development and systematic review methodology. All authors have reviewed the draft versions of the manuscript and approved the final version as submitted.

**Abstract**

Background: While the relations between sleep, cognition and behavior have been extensively studied in adolescents and school-aged children, very little attention has been given to preschoolers.

Objective: In this systematic review, our aim was to survey articles that address the link between sleep and both cognition and behavior in preschoolers (24 to 72 months old).

Methods: Four electronic databases were searched, namely Medline, Web of Science, PsycINFO and ERIC, completed by forward and backward citation search.

Results: Among the 1590 articles identified (minus duplicates), 26 met the inclusion criteria. Globally, studies with the largest sample sizes (N=13) found that a greater quantity or quality of sleep was associated with better behavioral and cognitive outcomes, while the others were less consistent.

Conclusion: Although the current literature seems to indicate that sleep is related to behavioral and cognitive development as early as preschool years, the strength of the associations (i.e. effect sizes) was relatively small. In addition to taking stock of the available data, this systematic review identifies potential sources of improvement for future research.

**Abbreviations**

BT Bedtime, CBCL Child Behavior Checklist, NSD Night Sleep Duration, NA Non-available, NS Non-significant, NW Night-waking, PBQ Preschool Behavior Questionnaire, PKBS Preschool and Kindergarten Behaviors Scale, SDQ Strength and Difficulty Questionnaire, SE Sleep efficiency, SOL sleep onset latency, SP sleep problems, SD total sleep duration, WT wake up time

**Summary**

This is the first systematic review of the literature on sleep and its relation to cognition and behavior in preschool-aged children. In comparison with the literature focused on school-aged children, knowledge involving preschoolers is rather sparse. A total of 26 studies were included in this review, which revealed a high degree of heterogeneity regarding the type and means of measuring sleep variables and behavioral and cognitive variables, as well as the statistical methods employed. Amongst the 13 articles with the largest sample sizes (top 50% of the included studies, 12 different populations), 12 found that a higher quantity or quality of sleep was associated with better behavioral and/or cognitive outcomes. Results point to an association between sleep, behavior and cognition as early as preschool years, but the strengths of associations reported in the articles were relatively small. Studies with a smaller sample size were less concordant. It is consistent with our findings that the strengths of association are small, and thus require large sample sizes to ensure statistical detection power. Different aspects of sleep were not associated with all cognitive or behavioral features in the same way, which underscores the need for specific measures rather than general ones such as "sleep problems" or "behavior problems" to be able to decipher the relationships. There is also a need for large longitudinal studies using objective measures and accounting for confounding factors. The child's genotype has recently been shown to have a moderating role in the association between sleep and behavior, and should be further explored.

**INTRODUCTION**

Debate continues over the precise function of sleep, especially with regard to its role in cognitive and behavioral capacities (Stickgold and Walker, 2005; Vertes and Siegel, 2005). In the literature, several hypotheses explain how sleep might be related to cognition and behavior. The "vigilance hypothesis", as described by Vriend et al. (2015), is perhaps the most instinctive one. It postulates that sleepiness is an intermediate link. A lack of sleep induces sleepiness – as shown in experimental studies on adolescents and school-aged children (Carskadon et al., 1981; Fallone et al., 2001) – which in turn has been associated with conduct problems, reduced attention, processing speed and working memory (Calhoun et al., 2012). This hypothesis proposes that sleep plays a passive role, while others suggest an active role such as "sleep-dependent memory consolidation", "synaptic homeostasis" and "amygdala medial-prefrontal cortex connectivity". The mechanism by which newly acquired knowledge becomes long-lasting memories, called "memory consolidation", is one of the cognitive processes for which the link to sleep has been the most studied (Stickgold, 2005). Memories are transferred from the hippocampal region to the neo-cortex through local synaptic consolidation processes and systems-level reorganization. There is increasing evidence that this process is sleep-dependent (Marshall et al., 2006). In parallel, the "synaptic homeostasis" hypothesis stipulates that to compensate for the persistent strengthening of synapses occurring during daytime activity, a synaptic downscaling takes place during slow wave sleep activity, allowing for better neuroplasticity (Tononi and Cirelli, 2006). The lack of synaptic plasticity, induced by inadequate sleep, could impair memory (Tononi and Cirelli, 2014), and play a role in depression (Liu et al., 2017) and maladaptive behavior (Kolb and Gibb, 2014). Another hypothesis is that sleep alters connectivity between the amygdala and the medial-prefrontal cortex, as observed in an experimental study conducted by Yoo et al. (2007). Since those structures seem to play an important role in the regulation of emotions (Phelps and LeDoux, 2005) and behavior (Amat et al., 2005), a lack of sleep could have a negative impact on these functions.

The relation between sleep, cognition and/or behavior has been extensively studied in adolescents and school-aged children (Astill et al., 2012; Vriend et al., 2015). According to the Astill et al's meta-analysis (2012), sleep duration in school-aged children is positively associated with executive functioning and multiple-domain cognitive function, and negatively associated with internalizing and externalizing behavior. No associations were found with sustained attention and memory. However, there has been little focus on preschoolers. This constitutes a gap in our knowledge, as sleep research on older children cannot be inferred to apply to preschoolers for numerous reasons. Sleep physiology, sleep need and sleep maturation evolve very rapidly in the first years of life, simultaneously to brain maturation (Louis et al., 1997; Olini and Huber, 2014). Additionally, the causes of insufficient quantity or quality of sleep can be quite different according to the child's age, and may give rise to distinct symptoms. While the brain and the circadian rhythm are still in maturation, inadequate sleep may disturb a child's development and thus be more likely to have long-term effects than when sleep disturbance occurs in adults, and different windows of exposure throughout childhood could lead to diverse consequences.

A systematic review of the literature on nap sleep among preschoolers was recently published (Thorpe et al., 2015), showing inconsistent results. Authors report this may be due to variation in age and in habitual napping status of the samples. In this article, we aim to survey all available publications focusing on the link between sleep, cognition and/or behavior in preschoolers, excluding studies focusing exclusively on naps. This systematic review includes studies published in English, conducted on children of the general population (non-clinical sample).

**METHODS**

The Cochrane handbook (Higgins and Green, 2008) and the Preferred Reporting Items for Systematic Reviews and Meta-analyses statement (Moher et al., 2009) (PRISMA guidelines) were followed to ensure, respectively, optimal methodology and optimal reporting (Supplementary data 1). This review has been declared in the PROSPERO database (number CRD42015029647).

**Eligibility criteria**

To be eligible, articles had to conform to a number of criteria. The exposure variable had to be a measure of night or total sleep duration, of difficulty initiating or maintaining sleep, or of sleep timing. The outcome variable needed to be a measure of cognition or behavior. The study population had to be a sample of the general population (not a clinical sample), with no selection according to the cognitive or behavioral status of the child or according to his or her sleep. The average age of the population had to be at least 2 and less than 6 years when the sleep and cognition measures were made. Moreover, only published original articles (no conference abstracts, reviews or case reports) written in English were eligible.

**Information source and search strategy**

The search strategy was implemented in two main steps. The first one consisted in a wide-ranging search of four electronic databases, namely Medline, Web of Science, PsycINFO and ERIC up to April 30, 2016. To ensure maximum sensitivity, no terms were used to exclude clinical studies, so that potential control groups could be included in our review. For all the databases, a search by title and abstract included the following words and their synonyms (the exact phrases are available in the supplementary data 2): *Sleep, Insomnia, night-waking, child, preschoolers, cognition, learning, teacher rating, performance, intelligence, IQ, memory, attention deficit, sustained attention, behavior, conduct, internalizing, externalizing.* For the Medline database, in addition to the title and abstract search, a second phrase was created to use Medical Subject Headings terms (hierarchically organized keywords also called MeSH terms): *Sleep, sleep disorders, child, preschool, child behavior, social behavior, behavioral symptoms, affect, temperament, intelligence, internal-external control, achievement, child behavior disorders, memory, verbal learning, cognition, psychomotor, performance.*

In the second step, we conducted backward and forward citation searches for all the articles retained after the initial screening of the title and abstract. Backward citation searching consists in identifying additional articles by reviewing the reference list of the selected articles, while forward citation searching consists in reviewing articles that cite the selected article. For the latter, we used the forward citation searching tools of both "PubMed" and "Web of Science".

**Article selection**

Two investigators selected the articles in two steps, and were blind to each other's selection throughout the whole process. The first selection was based on a screening process of the title and abstract and the second one was made after reading the full text.

**Data extraction and reporting**

The data extracted from the articles were the age of the child (average and range) at each measure, the recruitment date, the gender ratio, the sleep measure (the exposure variable), the cognitive and behavioral measures (the outcomes), the measurement tool used and by whom, the control variables included in the final model, the strength of the association (beta, correlation, odds-ratio, mean difference) and finally the significance of the association assessed by p-value. Data were extracted from the articles by the two investigators separately, and only concerned the analyses of interest for this review, i.e. sample size, study design, and control for confounding factors. Therefore, descriptions do not necessarily reflect the entire original article. The same remark applies to the risk of bias in assessment. If the article presented cross-sectional and longitudinal analyses, only the longitudinal ones were included. Similarly, if both unadjusted and multivariable analyses were available, the latter were reported. The child behavior checklist's total problems scale (Achenbach, 1991), but not its subscales, contains several questions regarding sleep difficulties. Thus, analyses using the total problems scale of the child behavior checklist were not reported.

**Risk of bias**

We assessed the risk of bias in each article using a 5-criteria scale, derived from the Newcastle Ottawa Quality Assessment Scale for Cohort studies (Wells et al., 2003). The first two criteria take into consideration classification bias, the third confounding bias, the fourth the lack of temporality and the fifth statistical detection power. Each of these elements accounted for a point on the risk of bias scale 1) the exposure (sleep variable) was not measured objectively; 2) the outcome (behavior or cognition) was assessed by someone who was not blind to the sleep status; 3) there were no statistical adjustments to take into account the main confounding factors (household income or parental education, child's sex, and child's age if age range > 6 months);  4) the study had a cross-sectional design (i.e. sleep was assessed at the same time as the outcome); 5) The sample size was less than 500 children. The maximum score of 5 indicates a higher risk of bias. The purpose of the scale is solely to give a general indication to readers of the level of confidence that can be given to the results. No articles were excluded based on this scale.

A same article may obtain a risk of bias range instead of a single score if it uses multiple methods. For example, in the same article, some cognitive measures might be assessed by a professional using a validated tool, while others might be assessed based on parents' perception.

**Data synthesis**

A systematic qualitative synthesis was performed. No reliable quantitative synthesis could be produced due to a great diversity of sleep, cognition and/or behavior measures, and to great variety in the data analysis methods, within a relatively low number of articles.

# RESULTS

## Selection and description of the studies included

The flow chart in Figure 1 summarizes the selection procedure, in accordance with the PRISMA guidelines (Moher et al., 2009). Out of the 1590 articles initially identified (minus duplicates), 1511 were excluded for not meeting inclusion criteria after title and abstract screening, and 53 were excluded after full text screening. We performed forward and backward citation searches on the articles that were selected for the full text assessment, which resulted in the inclusion of seven new articles. In total, 26 articles were included and are described in Table 1.

Out of the 26 selected articles, all reported observational studies. More than half were based on data collected in North American children (nine in the USA, six in Canada). The countries of origin of the children in the remaining articles were Australia (four), Europe (five in total, one in each of the following: Switzerland, Italy, Netherlands, Germany and Finland) and Japan (two).

As described in Figure 2, 14 articles (54%) had an estimated risk of bias of 4 or more out of 5. Eighteen (69%) articles reported a cross-sectional analysis, 13 (50%) had a population size below 500 and 23 (88%) used subjective measures of sleep. Regarding the outcome, 14 articles (54%) used measures assessed by someone who was not blind to the child sleep status and three had a combination of blind and non-blind assessors. Twenty-one articles (73%) did not meet minimum adjustment criteria (namely any socio-economic factors, child's sex, and child's age if age range > 6 months).

Figure 3 shows the number of articles per exposure and per outcome. In total, 24 different questionnaires for sleep, behavior and cognition were used, and are reported in supplementary data 3. Regarding exposure, over half of the articles focused on the child's sleep duration, 13 for night sleep duration (NSD) and six for total sleep duration (TSD). A few examined indicators of difficulty initiating and maintaining sleep such as night-waking (NW) (n=8), sleep onset latency (SOL) (n=5), sleep efficiency (SE) (n=2) and insomnia (I) (n=1). Sleep problems (SP) (n=6) were studied with varying definitions. In even fewer cases, bedtime (BT) (n=3) and wake-up time (WT) (n=1) were also investigated. Regarding the outcome, 23 of the publications raised the question of an association between sleep and behavior. Cognition was less frequently studied (n=7) with a notable focus on language skills (five out of seven articles).

**Sleep and behavior**

A summary of the associations found between sleep and behavior is presented in Table 2.

*Sleep and externalizing behavior*

*Aggressiveness and conduct problems*

Aggressive behavior has been found to be positively associated with insomnia (Armstrong et al., 2014), sleep problems (Hall et al., 2007; Hatzinger et al., 2010) and bedtime (Komada et al., 2011). Results showing night sleep duration were divergent, and no associations were found with sleep onset latency (Hatzinger et al., 2010) nor sleep efficiency (Hatzinger et al., 2010). More specifically, in the Armstrong et al. study (2014) of 396 children aged 54 months, those with insomnia had a higher mean score on the hostile-aggressive scale of the Preschool Behavior Questionnaire (PBQ) (0.81± 0.26). Hall et al. (2007) found that sleep problems at age 3 accounted for 5.1% of the variance on the Child Behavior Check List (CBCL) scale of aggressive behavior at age 4 (n=1317). The correlation between aggressive behavior (measured by CBCL) and bedtime on week days was r=0.11 (p<0.01) at ages 2 to 3 and 4 to 5 in the Komada et al. study (2011). Furthermore, Komada et al. (2011) showed that shorter night sleep duration was associated with more aggressive behavior in children between the age of 24 to 36 months (N=905), but not for those between the ages of 48 to 60 months (N=841). Hatzinger et al. (2010) also found a lack of association in 84 children aged 59 months. These results seem to indicate that shorter night sleep duration is associated with more aggressive behavior in younger children only. However Scharf et al. (2013) did find a positive association in their sample of 8950 children aged 48 months. Children who slept less than 9.44 hours per night had an increased risk of being frequently aggressive (measured by Preschool and Kindergarten Behaviors Scale or PKBS) with an OR of 1.81 CI 95% [1.36-2.41]. Thus, we could not determine whether the inconsistencies in the results were due to differences in sample size or in sample age.

There were disparities in the results on conduct problems. These might be caused by differences in sample size and thus detection power. Wada et al. (2013) and Hatzinger et al. (2010) found in 431 and 84 children, respectively, that none of the studied sleep exposures - namely night and total sleep duration, sleep onset latency, night-waking sleep efficiency, sleep problems, bedtime and wake-up time - were associated with conduct problems. In studies with larger sample size, children who had more night-waking (Hiscock et al., 2007; Lehmkuhl et al., 2008), longer sleep onset latency (Hiscock et al., 2007; Lehmkuhl et al., 2008), and more sleep problems (Hiscock et al., 2007; Quach et al., 2012), had a higher risk of conduct problems.

More specifically, in the Hiscock et al. study (2007) of 4983 children aged 57 months, those who woke at night or had difficulty getting to sleep 4 nights per week or more, had a higher score on the SDQ conduct problem scale of 0.6 CI 95% [0.5-0.8] and 1.0 points CI 95% [0.8-1.2], respectively. Compared to the children with no sleep problems, those with a moderate to severe sleep problem, as declared by parents, had a mean difference of 1.1 points CI 95% [0.9-1.3] on that same scale. Although the definition of sleep problems was different in Quach et al. (2012), they found very similar results in 1512 children aged 68 months, with a mean difference in the conduct problems score of 1.0 points CI 95% [0.7-1.2] between no sleep problems and moderate to severe sleep problems. We note that the only study using objective measures of sleep found no significant association (Hatzinger et al., 2010).

*Attention and hyperactivity problems*

Some sleep parameters, such as later bedtime (Komada et al., 2011), and global sleep problems (O'Callaghan et al., 2010), were positively associated with attention problems. Regarding the strength of these associations, the correlation between the concurrent score of attention problems (measured by the CBCL) and bedtime was r=0.10 (p<0.01) in Komada et al. study (2011). Children who had sleep problems occurring "often" between the age of 2 and 4 years had increased risk of persistent attention problems (above the 90[th] percentile on the CBCL attention scale at age 5 and 14) with an adjusted OR of 3.84 CI 95% [2.23-6.64] for boys, and 4.42 CI 95% [2.27-8.63] for girls in the O'Callaghan et al. study (2010). No links were found with night-waking (Hall et al., 2012), sleep onset latency (Vaughn et al., 2015), sleep efficiency (Hatzinger et al., 2010) nor wake-up time variability (Vaughn et al., 2015). Regarding night sleep duration, Paavonen et al. (2009) described a positive association when attention problems were assessed by parents, but not when assessed by teachers. These results could be indicative of a potential bias when the person reporting the behavioral measure is not blind to the sleep status, although the authors also point out other plausible explanations for the discrepancy. For instance, the teacher response rate was low, and not missing at random since it was lower for children with greater behavioral difficulties according to parents. Four other studies (Komada et al., 2011; Lam et al., 2011; Touchette et al., 2007; Vaughn et al., 2015) reported a lack of association regardless of who evaluated the child's attention.

Dissimilarities in results were observed for every studied association between a sleep factor and hyperactivity problems.

### Sleep and internalizing behavior

*Anxious, depressed*

In the Jansen et al. study (2011) of 4782 children aged 24 months, those who slept less than 12.5 hours per day at age 2 years, compared to those who slept more than 13.5 hours, had an increased risk of anxiety or depressive symptoms at age 3 (defined as being above the $80^{th}$ percentile on the CBCL anxious/depressed symptoms scale) with an adjusted OR of 1.47 CI 95% [1.20-1.79]. Results were similar when looking at the child's night-waking, with an adjusted OR of 1.32 CI 95% [1.14-1.54] for children who woke once or twice per night on average compared to those who never woke. Similarly, Zaidman et al. (2015) found in 1487 children aged 29 months that those who woke 20 minutes or more per night, compared to those who did not wake, had higher anxiety and depression on an adapted scale of the PBQ. In Komada et al. (2011), the correlation between the CBCL anxious/depressed symptoms scale and bedtime was r=0.09 (p<0.01). The associations were not significant when studying night sleep duration (Komada et al., 2011) and insomnia (Armstrong et al., 2014).

*Emotional symptoms*

Studies with the lowest sample sizes reported non-significant results for all sleep measures, namely night sleep duration (Hatzinger et al., 2010; Wada et al., 2013), total sleep duration (Wada et al., 2013), night-waking (Hall et al., 2012; Hatzinger et al., 2010; Wada et al., 2013), sleep onset latency (Hatzinger et al., 2010; Wada et al., 2013), sleep problems (Hatzinger et al., 2010), bedtime (Wada et al., 2013) and sleep efficiency (Hatzinger et al., 2010) in children aged around 60 months. In contrast, those with larger sample sizes reported significant results for all reported measures. Night-waking (Hiscock et al., 2007; Lehmkuhl et al., 2008), sleep onset latency (Hiscock et al., 2007; Lehmkuhl et al., 2008) and sleep problems (Hiscock et al., 2007; Quach et al., 2012) were positively and significantly associated with emotional problems in children aged between 57 and 68 months. Although the sample size might not be the sole factor explaining the differences in results, it seems that lack of statistical power was common in this matter. Other internalizing behaviors were seldom investigated, thereby limiting a comprehensive synthesis.

*Sleep, social behavior and peer relations*

Children with shorter night sleep (Vaughn et al., 2015; Wada et al., 2013) or who globally had more sleep problems (Hiscock et al., 2007; Quach et al., 2012), were reported as showing less prosocial behavior. However, no significant association was found, either with total sleep duration, bedtime nor sleep efficiency. Lehmkuhl et al. (2008) surprisingly found among 1388 children aged 66 months that children with longer sleep onset latency displayed more prosocial behavior (N=1338) where Hiscock et al. (2007) found the opposite in 4983 children. The main difference in method in the two articles was that Lehmkuhl et al. (2008) reported simple bivariate analyses while Hiscock et al. (2007) adjusted for age, gender and household income. Two other articles with smaller sample sizes (in 62 and 437 children, respectively) found no significant association (Vaughn et al., 2015; Wada et al., 2013). In each article, very similar results were observed for the outcome denoted by acceptance by peers.

**Sleep and Cognition**

Even fewer studies have investigated the link between sleep and cognition in preschoolers (N=7). They are presented in Table 3. Out of the four articles studying the association between night sleep duration and receptive vocabulary capacities, three found a positive association. In the Lam et al. (2011) cross-sectional study among 59 children aged 36 to 60 months, actigraphy-measured night sleep was positively correlated with a better score on the PPVT-IV receptive vocabulary scale (age adjusted r=0.29, p=0.03). Also using actigraphy and the PPVT-IV in a cross-sectional study, Vaughn et al. (2015) found in 62 children aged 50 months a correlation of r=0.45 (p<0.01) after adjusting for age, sex and ethnicity. In their longitudinal study of 1492 children, Touchette et al. (2007) described four night-sleep patterns from age 2.5 to 6: 11-hour persistent, 10-hour persistent, short increasing, and short persistent. Compared to children who slept 11-hour persistently, those who had short persistent duration were at higher risk of a low PPVT-measured receptive vocabulary score (p=0.001), but no significant association was found with other sleep patterns. Dionne et al. (2011) found that parental reports of night sleep duration at 30 months was not associated with concurrent receptive vocabulary (assessed by the MCDI), nor predictive of the 60 months receptive vocabulary (assessed by the PPVT) in 1029 children. Hiscock et al. (2007) found that sleep problems were associated with literacy and numeracy but not with receptive vocabulary. No associations were found between any cognitive outcome and total sleep duration, night-waking, sleep onset latency or sleep efficiency.

**DISCUSSION**

Amongst the 13 articles with the largest sample sizes (top 50% of the selected studies, 12 different populations), 12 found that a higher quantity or quality of sleep was associated with better behavioral and/or cognitive outcomes. Results point to an association between sleep, behavior and cognition as early as in preschool years. Studies with a smaller sample size were less consistent. The strengths of associations reported in the articles were relatively small, which explains the need for a large sample size to find consistent results. The studies were heterogeneous in many regards: the type and means of measures for the sleep variables differed but also the behavioral and cognitive variables, as well as the statistical methods employed. Results differed according to the specific exposure and outcome considered, as well as the method employed, but too few studies were performed to fully understand specific associations.

In comparison with the literature based on school-aged children, knowledge involving preschoolers is rather sparse. Astill et al. (2012) conducted a systematic review using similar selection criteria and found over 80 studies examining the association between sleep and cognition or behavior in children between the ages of 5 and 12. According to their meta-analysis, the association between sleep duration and both cognition (r=0.08) and behavior (r=0.09) in school-aged children was small but significant. Our findings suggest these associations could be found as early as preschool-years.

**Quality of included studies**

Unlike what is seen in the day-time sleep literature (Thorpe et al., 2015), no experimental studies have been performed on preschoolers' night sleep. Although they provide better strength of evidence than observational studies, experiments in very young children raise ethical concerns and problems of parental acceptance, which explains their absence. Furthermore, experimental studies would not make it possible to observe the effect of long-term suboptimal sleep.

Most of the studies included do not use objective sleep measures, only three used actigraphy, and none used polysomnography, the gold standard in sleep measures. Objective measures can be expensive to record and to analyze especially for larger samples. It can also be difficult to obtain access to polysomnographic equipment for research purposes, which no doubt further limits its use. Actigraphy is a non-invasive objective method which is becoming more and more accessible, we can thus expect a great increase in objective sleep measures in future research. However, some concerns have been raised regarding the validity of this measure in preschool-aged children. When compared to polysomnography, correlations were above 0.80 for sleep latency, sleep duration and sleep efficiency but below 0.40 for the number of awakenings (Bélanger et al., 2013). Sitnick et al. (2008) suggest videotaping as a more reliable alternative.

Presumably for similar reasons, behavioral and cognitive data were often collected using questionnaires completed by the parents, who are not blind to the child's sleep status. Evaluation by an external investigator, ideally a psychologist, limits the risk of a bias in the outcome estimators. Failure to consider confounding factors was another common risk of bias. This shortcoming limits the interpretation of the results since it becomes impossible to determine whether the associations found, or the lack of associations, are dependent on other factors. This risk of bias can be easily reduced by statistical adjustment, if data on confounders are collected. Cross-sectional designs also make the interpretation of the results quite difficult, as they prevent determining whether the sleep outcome occurred before or after the behavior or cognitive impairment. The chronology of events is especially important in this field of research, as there are bidirectional associations between sleep, and both cognition and behavior (Touchette et al., 2009).

**Reporting bias**

The growing literature regarding publication bias and selective reporting bias suggests they are widespread (Dwan et al., 2008). It is likely that they are even more frequent in reviews including observational studies, since their registration is not required (unlike clinical trials). Non-significant results that are published are also less frequently cited in other publications, reducing the likelihood of being identified. Reporting biases are complex to observe but we did find dissimilarities in several studies between the measures used in the analyses and those apparently available. It is possible that authors solely reported associations with significant results, omitting other measures. One way to counter this ubiquitous problem would be to plan the study of exposures and outcomes ahead of analyses.

The aim of this review is to present available peer reviewed literature – the gray literature was therefore not searched. While this allows a higher quality of included articles, it may reinforce the impact of publication bias.

**Variability in definitions and means of measure**

The variability in definitions of outcomes and exposure limits the present understanding of the associations between sleep, behavior and cognition. For example, in the six articles studying "sleep problems" (Hall et al., 2007; Hatzinger et al., 2010; Hiscock et al., 2007; O'Callaghan et al., 2010; Quach et al., 2012; Troxel et al., 2013), none had similar definitions. While most included a notion of difficulty initiating or maintaining sleep, some also included sleep habits, some mixed dyssomnia and parasomnia and others included only one simple question assessing parents' perception of the child's sleep. The means to measure the outcome also varied greatly, with 13 different tools used to assess behavior and eight for cognition (described in supplementary data 3). The quality and comparability could easily be enhanced if the studies focused on specific aspects of both exposure and outcome, especially since one cannot assume that different specific sleep exposures are related to all cognitive or behavioral features in the same way.

**Future directions**

This study highlights the need to reduce publishing bias as well as bias within studies. For the latter, recommendations are to explore the association between specific sleep, behavioral and cognitive measures through longitudinal studies, to seek sufficient statistical power and use objective sleep measures, as well as having the behavioral or cognitive outcome assessed by an investigator blind of the sleep status, and report all available analyses. Reporting bias and confounding bias – as well as vague results arising from the study of nonspecific sleep, cognitive and/or behavioral measures – can be easily reduced if taken into account during the planning process. Reporting the children's age precisely and limiting the age range in a same analysis could also improve interpretation.

It is of note that some recent studies have focused on less typical sleep aspects such as daily variations (Spruyt et al., 2016) and chronotype (Doi et al., 2015). Others have explored sleep hygiene, such as bedtime routine (Mindell et al., 2015), showing the importance of educating parents on the matter. Mindell et al. (1994) found in their study that treatment of sleep disturbance improved day-time behavior. New findings also report a differential role of sleep according to the child's genotype (Bouvette-Turcot et al., 2015). Our present understanding in this field could be greatly improved by these innovative research approaches.

A recent systematic review and meta-analysis found behavioral interventions to be efficient in reducing sleep problems in children (Meltzer et al., 2014). The interventions for preschool-aged children included sleep education, graduated extinction, structured bedtime routine and sleep programs. According to their meta-analyses, the standard mean deviation between the intervention and control group was 0.33 (0.48–0.18), and the mean night waking frequency in the intervention groups was 0.26 standard deviations lower (0.35–0.17). It would be interesting to investigate if such interventions also improve cognition and behavior

**CONCLUSION**

In this systematic review, we took stock of the available data on the question of the association between sleep, behavior and cognition in preschoolers, and suggested ways to improve future research on the subject. The current literature seems to indicate that sleep is related to behavioral and cognitive development as early as in preschool years.

**ACKNOWLEDGEMENTS**

We thank Dr Frank Ramus, who shared his expertise on child cognitive and behavioral development. We also thank Pr Isabelle Boutron, co-convenor of the Bias Methods group of the Cochrane Collaboration, for her guidance in the systematic review method.

**Figures**

**Figure 1. Systematic review flow chart, following PRISMA guidelines** (Moher et al., 2009)

**Figure 2. Quality of included studies.** a) Article distribution according to the five quality criteria, b) Paper distribution according to the total risk of bias (sum of the five quality criteria)

**Figure 3. Number of articles per exposure and per outcome** (NSD night sleep duration, SP sleep problems, NW night-waking, TSD total sleep duration, SOL sleep onset latency, BT bedtime, SE sleep efficiency, WT wake up time)

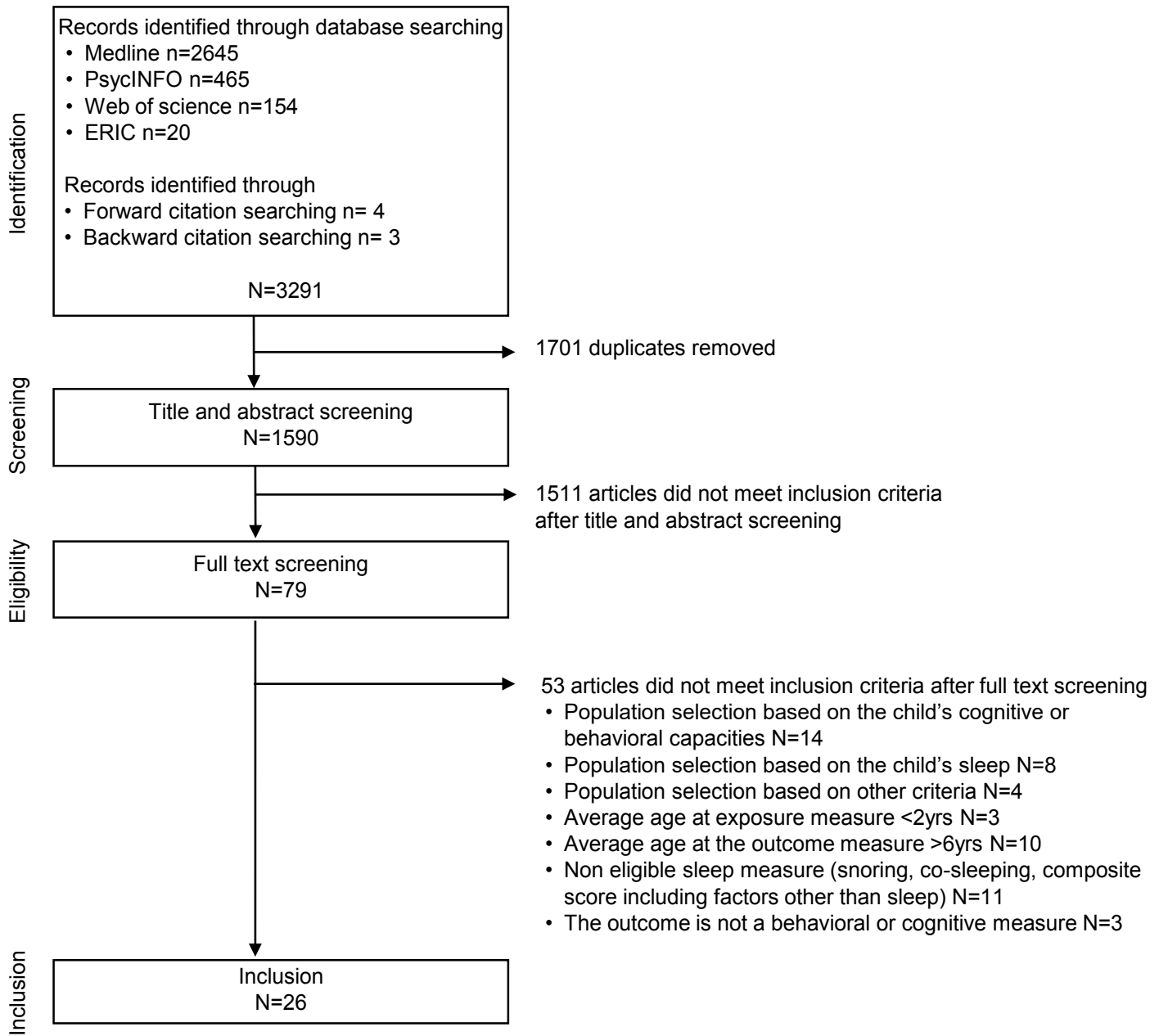

**a) Article distribution according to the 5 quality criteria**

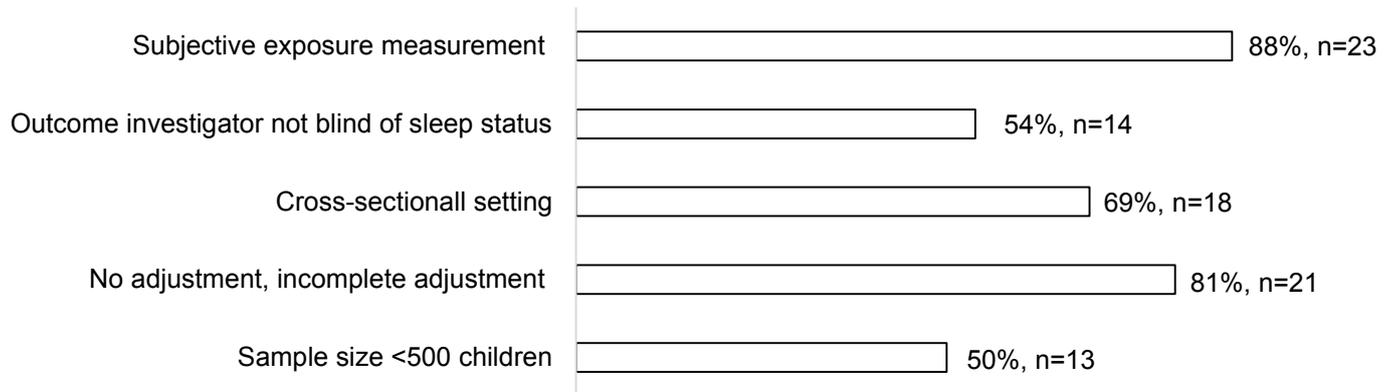

**b) Article distribution according to the total risk of bias (sum of the 5 criteria)**

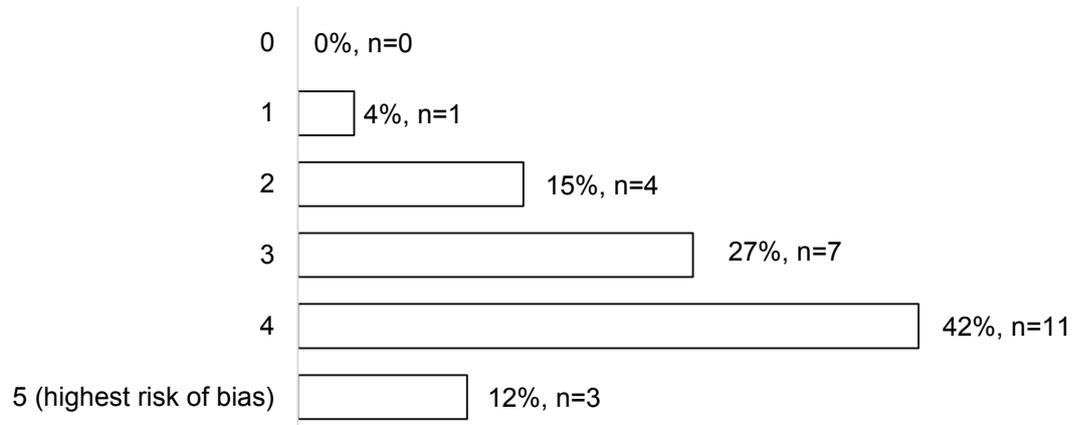

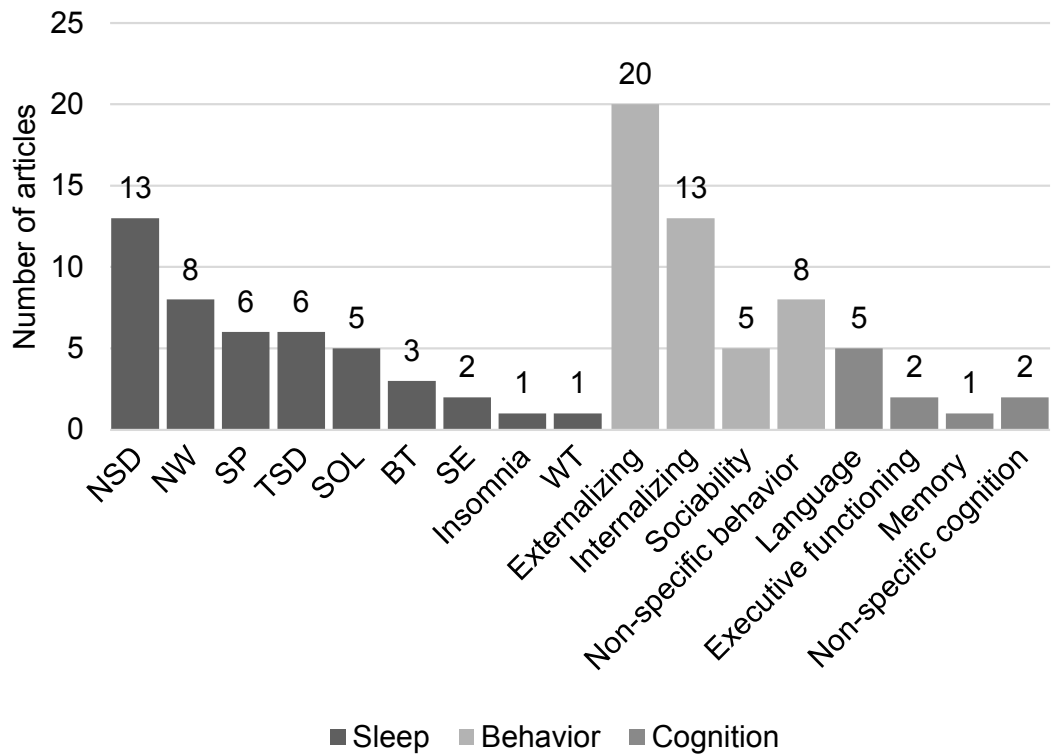

**Table 1. Description of the 26 included studies**

| Author, year | N | Country | Objective sleep measure | Design | Age in months (±SD or range) At exposure | Age in months (±SD or range) At outcome if different | Control for confounding factors [a] | Risk of bias [b] |
|---|---|---|---|---|---|---|---|---|
| Armstrong et al. (2014) | 396 | USA | No | CS [c] | 54(±NA) | - | No [c] | 5 [c] |
| Bates et al. (2002) | 184 | USA | No | CS | 58.8(±6.5) | - | Partial | 4 |
| Bouvette-Turcot et al. (2015) | 209 | CAN | No | L | 12 to 36 [d] | 36(±NA) | Yes | 3 |
| Bruni et al. (2000) | 194 | ITA | No | CS | 27(22-38) | - | No | 5 |
| Dionne et al. (2011) | 1029 | CAN | No | L | 31(±0.8) | 31(±0.8) & 63(±3.0) | No [c] | 2 [c] |
| Hall et al.(2007) | 1317 | AUS | No | L | 36(±NA) [c] | 48(±NA) | Partial | 3 |
| Hall et al. (2012) | 58 | CAN | No | CS | 24.7(±7.0) | - | No | 4-5 |
| Hatzinger et al. (2010) | 82 | CHE | Yes | CS | 58.9(±5.8) | - | No | 3 |
| Hiscock et al. (2007) | 4983 | AUS | No | CS | 56.9(51-67) | - | Both | 2-4 |
| Jansen et al. (2011) | 4782 | NLD | No | L | 24(±NA) | 36(±NA) | Yes | 2 |
| Jung et al. (2009) | 67 | USA | No | L | 42.1(±3.3) | 42 to 65 [d] | No | 3 |
| Komada et al. (2011) | 1746 | JPN | No | CS | (24-36) & (48-60) | - | No [c] | 4 [c] |
| Lam et al. (2011) | 59 | USA | Yes | CS | 51.6(36-60) | - | Partial | 3 |
| Lehmkuhl et al. (2008) | 1388 | DEU | No | CS | 66.2(±NA) | - | No | 4 |
| Nathanson and Fries (2014) | 107 | USA | No | CS | 53.4(±8.7) | - | Partial | 4 |
| O'Callaghan et al. (2010) | 4204 | AUS | No | CS [c] | (24-48) [e] | 60(±NA) | Partial [c] | 4 [c] |
| Paavonen et al. (2009) | 297 | FIN | No | CS | (60-72) | - | Partial | 4 |
| Quach et al. (2012) | 1512 | AUS | No | CS | 68.4(±4.8) | - | Partial | 4 |
| Scharf et al. (2013) | 8950 | USA | No | CS | 48(±NA) | - | Partial | 4 |
| Touchette et al. (2007) [f] | 1492 | CAN | No | L | 30 to 72 [d] | 61(±3.6) [c] | Yes | 1 |
| Touchette et al. (2009) [f] | 2057 | CAN | No | L | 18 to 60 [d] | - | No [c] | 3 [c] |
| Troxel et al. (2013) | 776 | USA | No | L | 24(±NA) & 36(±NA) | 54(±NA) | Partial | 2 |
| Vaughn et al. (2015) | 62 | USA | Yes | CS | 49.8(±7.4) | - | Partial | 3 |
| Wada et al. (2013) | 437 | JPN | No | CS | 61.4(±10.8) | - | Yes | 4 |
| Weissbluth (1984) | 60 | USA | No | CS | 36.1(36-38) | - | No | 5 |
| Zaidman-Zait and Hall (2015) | 1487 | CAN | No | CS [c] | 29(±NA) | - | Partial | 4 |

[a] Adjustment on sex, socio economic factors, and age when sample age range >6 months
[b] From 0 to 5, a higher score indicating a higher risk of bias
[c] Concern the analyses of interest to the review
[d] Repeated measures
[e] Measure assessed at 60 months (retrospective)
[f] Same study sample

AUS = Australia, CAN = Canada, CHE = Switzerland, DEU = Germany, FIN = Finland, ITA = Italy, JPN = Japan, NLD = Netherlands, USA =United States of America, L longitudinal analysis, CS cross-sectional analysis, NA Non-available data

**Table 2 Associations between sleep and behavior in preschoolers**

| | TSD | NSD | I | NW | SOL | SE | SP | BT | WT |
|---|---|---|---|---|---|---|---|---|---|
| **EXTERNALIZING BEHAVIOR** | | | | | | | | | |
| **Aggressiveness** | | | | | | | | | |
| Armstrong et al. (2014) | | | +** | | | | | | |
| Hall et al. (2007) | | | | | | | +*** | | |
| Hall et al. (2012) | | | | NS | | | | | |
| Hatzinger et al. (2010) | | | | | | | +** | | |
| Komada et al. (2011), 24-36 mo | | -** | | | | | | +** | |
| Komada et al. (2011), 48-60 mo | | NS | | | | | | +** | |
| Scharf et al. (2013) | | -*** | | | | | | | |
| Zaidman-Zait and Hall (2015) | | | | +* | | | | | |
| **Anger** | | | | | | | | | |
| Scharf et al. (2013) | | -** | | | | | | | |
| **Attention problems** | | | | | | | | | |
| Hall et al. (2012) | | | | NS | | | | | |
| Komada et al. (2011), 24-60 mo | | NS | | | | | | +** | |
| Lam et al. (2011) | | NS | | | | | | | |
| O'Callaghan et al. (2010) | | | | | | | +*** | | |
| Paavonen et al. (2009), parents report | | -** | | | | | | | |
| Paavonen et al. (2009), teacher report | | NS | | | | | | | |
| Touchette et al. (2007) | | NS | | | | | | | |
| Vaughn et al. (2015) | | NS | | | NS | NS | | | |
| **Conduct problems** | | | | | | | | | |
| Hatzinger et al. (2010), boys | | NS | | +* | NS | NS | | | |
| Hatzinger et al. (2010), girls | | NS | | NS | NS | NS | | | |
| Hiscock et al. (2007) | | | | +*** | +*** | | +*** | | |
| Lehmkuhl et al. (2008) | | | | +** | +** | | | | |
| Quach et al. (2012) | | | | | | | +*** | | |
| Wada et al. (2013) | NS | -* | | NS | NS | | | NS | NS |
| Scharf et al. (2013) (Annoying behavior) | | NS | | | | | | | |
| Scharf et al. (2013) (*Tantrums*) | | -* | | | | | | | |
| **Hyperactivity** | | | | | | | | | |
| Armstrong et al. (2014) | | | +** | | | | | | |
| Hatzinger et al. (2010), boys | | NS | | +* | NS | NS | | | |
| Hatzinger et al. (2010), girls | | NS | | NS | NS | NS | | | |
| Lam et al. (2011) | | NS | | | | | | | |
| Lehmkuhl et al. (2008) | | | | NS | +** | | | | |
| Quach et al. (2012) | | | | | | | +*** | | |
| Scharf et al. (2013) | | -** | | | | | | | |
| Touchette et al. (2009) | | -*** | | | | | | | |
| Wada et al. (2013) | NS | NS | | +* | NS | | | NS | +* |
| Zaidman-Zait and Hall (2015) | | | | +*** | | | | | |
| *Hyperactivity and impulsivity* | | | | | | | | | |
| Touchette et al. (2007) | | -*** a | | | | | | | |
| *Hyperactivity and attention* | | | | | | | | | |
| Hiscock et al. (2007) | | | | +*** | +*** | | +*** | | |
| **Impulsivity** | | | | | | | | | |
| Scharf et al. (2013) | | -*** | | | | | | | |
| **Opposition** | | | | | | | | | |
| Zaidman-Zait and Hall (2015) | | | | +** | | | | | |
| **Non-specific externalizing behavior** | | | | | | | | | |
| Bruni et al. (2000) | | | | +* | | | | | |
| Hall et al. (2012) | | | | NS | | | | | |
| Paavonen et al. (2009), parents report | | NS | | | | | | | |

**Table 2 continued)**

| | TSD | NSD | I | NW | SOL | SE | SP | BT | WT |
|---|---|---|---|---|---|---|---|---|---|
| Paavonen et al. (2009), teacher report | | NS | | | | | | | |
| Scharf et al. (2013) | | -*** | | | | | | | |
| Troxel et al. (2013) | | | | | | | +* | | |
| Weissbluth (1984) | NS | NS | | | | | | | |
| **INTERNALIZING BEHAVIOR** | | | | | | | | | |
| **Anxious, depressed** | | | | | | | | | |
| Armstrong et al. (2014) | | | NS | | | | | | |
| Hall et al. (2012) | | | | NS | | | | | |
| Jansen et al. (2011) | -* | | | +* | | | | | |
| Komada et al. (2011), 24-36 mo | | NS | | | | | | +** | |
| Komada et al. (2011), 48-60 mo | | NS | | | | | | +** | |
| Zaidman-Zait and Hall (2015) | | | | +* | | | | | |
| **Separation anxiety** | | | | | | | | | |
| Zaidman-Zait and Hall (2015) | | | | +*** | | | | | |
| **Emotional symptoms** | | | | | | | | | |
| Hall et al. (2012) | | | | NS | | | | | |
| Hatzinger et al. (2010) | | NS | | NS | NS | NS | NS | | |
| Hiscock et al. (2007) | | | | +*** | +*** | | +*** | | |
| Lehmkuhl et al. (2008) | | | | +** | +** | | | | |
| Quach et al. (2012) | | | | | | | +*** | | |
| Wada et al. (2013) | NS | NS | | NS | NS | | | +* | NS |
| **Withdrawn, shyness** | | | | | | | | | |
| Hall et al. (2012) | | | | NS | | | | | |
| Zaidman-Zait and Hall (2015) | | | | +** | | | | | |
| **Somatization** | | | | | | | | | |
| Bruni et al. (2000) | | | | +** | | | | | |
| Hall et al. (2012) | | | | NS | | | | | |
| **Non-specific internalizing behavior** | | | | | | | | | |
| Bruni et al. (2000) | | | | NS | | | | | |
| Hall et al. (2012) | | | | NS | | | | | |
| Paavonen et al. (2009), parents report | | -* | | | | | | | |
| Paavonen et al. (2009), teacher report | | NS | | | | | | | |
| Troxel et al. (2013) | | | | | | | +** | | |
| **SOCIABILITY** | | | | | | | | | |
| **Prosocial behavior** | | | | | | | | | |
| Hiscock et al. (2007) | | | | -*** | -*** | | -*** | | |
| Lehmkuhl et al. (2008) | | | | NS | +** | | | | |
| Quach et al. (2012) | | | | | | | -*** | | |
| Vaughn et al. (2015) | | +* | | | NS | NS | | | |
| Wada et al. (2013) | NS | NS | | NS | NS | | | -* | -** |
| **Peer relation problems** | | | | | | | | | |
| Hiscock et al. (2007) | | | | +** | +*** | | +*** | | |
| Lehmkuhl et al. (2008) | | | | NS | NS | | | | |
| Quach et al. (2012) | | | | | | | +*** | | |
| Vaughn et al. (2015) | | -** | | | NS | NS | | | |
| Wada et al. (2013) | NS | NS | | NS | NS | | | NS | +** |
| **NON-SPECIFIC BEHAVIOR PROBLEMS** | | | | | | | | | |
| Armstrong et al. (2014) | | | +* | | | | | | |
| Bates et al. (2002) | NS | NS | | | | | | NS | |
| Bouvette-Turcot et al. (2015) ≥1 copy of the 5-HTTLRPR [b] short allele | -*** | | | | | | | | |
| Bouvette-Turcot et al. (2015) no copy of the 5-HTTLRPR [b] short allele | NS | | | | | | | | |

**Table 2 continued)**

| | TSD | NSD | I | NW | SOL | SE | SP | BT | WT |
|---|---|---|---|---|---|---|---|---|---|
| Hiscock et al. (2007) | | | | +*** | +*** | | +*** | | |
| Lehmkuhl et al. (2008) | | | | +** | +** | | | | |
| Paavonen et al. (2009), teacher report | | NS | | | | | | | |
| Quach et al. (2012) | | | | | | | +*** | | |
| Wada et al. (2013) | NS | NS | | +* | NS | | | +* | +* |

Note: +, - and NS indicate a positive, negative and non-significant statistical association
Abbreviations stand for: TSD total sleep duration, NSD night sleep duration, I insomnia, NW night-waking, SOL sleep onset latency, SE sleep efficiency, SP sleep problems, BT bed time, WT wake-up time.
* p<0.05, ** p≤0.01, ***p≤0.001
[a] When compared to the 11-hour persistent sleep duration trajectory, the short increasing duration trajectory is a risk factor p=0.001, but no significant association was found with other categories
[b] Serotonin-transporter-linked polymorphic region of the serotonin transporter gene (SLC6A4)

**Table 3 Associations between sleep and cognition in preschoolers.**

|  |  | TSD | NSD | NW | SOL | SE | SP |
|---|---|---|---|---|---|---|---|
| **Executive function** | | | | | | | |
| Lam et al. (2011) | *ACPT-P Omission error* | | NS | | | | |
| Lam et al. (2011) | *ACPT-P Commission error* | | -* | | | | |
| Lam et al. (2011) | *ACPT-P Mean response time* | | NS | | | | |
| Lam et al. (2011) | *ACPT-P Variability* | | NS | | | | |
| Nathanson and Fries (2014) | | NS | | | | | |
| **Memory** | | | | | | | |
| Lam et al. (2011) | | | NS | | | | |
| **Language** | | | | | | | |
| Dionne et al. (2011) | *Receptive Vocabulary* | | NS | | | | |
| Hiscock et al. (2007) | *Receptive Vocabulary* | | | NS | NS | | NS |
| Hiscock et al. (2007) | *Literacy and numeracy* | | | NS | NS | | +*** |
| Lam et al. (2011) | *Receptive Vocabulary* | | +* | | | | |
| Vaughn et al. (2015) | *Receptive Vocabulary* | | +** | | NS | NS | |
| Touchette et al. (2007) | *Receptive Vocabulary* | | +*** [a] | | | | |
| **Non-specific cognition** | | | | | | | |
| Jung et al. (2009) | *General conceptual ability* | | +* | | | | |
| Touchette et al. (2007) | *Non-verbal abilities* | | +*** [b] | | | | |

Note: +, - and NS indicate a positive, negative and non-significant statistical association
Abbreviations stand for: TSD total sleep duration, NSD night sleep duration, NW night-waking, SOL sleep onset latency, SE sleep efficiency, SP sleep problems.
* p<0.05, ** p≤0.01, ***p≤0.001
ACPT-P Auditory continuous performance test for preschoolers (Mahone et al. (2001)
[a] with the reference being children who slept 11-hour persistently, children who had short persistent duration were at higher risk (p=0.001), but no significant association was found with other categories
[b] with the reference being children who slept 11-hour persistently, children who had short increasing duration were at higher risk (p=0.001), but no significant association was found with other categories